\documentclass[sigconf]{acmart}
\AtBeginDocument{%
  }

\copyrightyear{2026}
\acmYear{2026}
\setcopyright{cc}
\setcctype{by}
\acmConference[CHI '26]{Proceedings of the 2026 CHI Conference on Human Factors in Computing Systems}{April 13--17, 2026}{Barcelona, Spain}
\acmBooktitle{Proceedings of the 2026 CHI Conference on Human Factors in Computing Systems (CHI '26), April 13--17, 2026, Barcelona, Spain}
\acmPrice{}
\acmDOI{10.1145/3772318.3791557}
\acmISBN{979-8-4007-2278-3/2026/04}

\begin{document}

\title[Exploring the Impacts of Background Noise on Auditory Stimuli of Audio-Visual eHMIs]{Exploring the Impacts of Background Noise on Auditory Stimuli of Audio-Visual eHMIs for Hearing, Deaf, and Hard-of-Hearing People}

\author{Wenge Xu}
\orcid{0000-0001-7227-7437}
\email{Wenge.Xu@bcu.ac.uk}
\affiliation{%
 \institution{Birmingham City University}
 \city{Birmingham}
 \country{United Kingdom}
}

\author{Foroogh Hajiseyedjavadi}
\orcid{0000-0003-3448-1239}
\email{Foroogh.Hajiseyedjavadi@bcu.ac.uk}
\affiliation{%
 \institution{Birmingham City University}
 \city{Birmingham}
 \country{United Kingdom}
}

\author{Debargha Dey}
\email{d.dey@tue.nl}
\orcid{0000-0001-9266-0126}
\affiliation{%
  \institution{Eindhoven University of Technology}
  \city{Eindhoven}
  \country{The Netherlands}
 }

\author{Tram Thi Minh Tran}
\email{tram.tran@sydney.edu.au}
\orcid{0000-0002-4958-2465}
\affiliation{Design Lab, Sydney School of Architecture, Design and Planning
  \institution{The University of Sydney}
  \city{Sydney}
  \state{NSW}
  \country{Australia}
}

\author{Mark Colley}
\orcid{0000-0001-5207-5029}
\email{m.colley@ucl.ac.uk}
\affiliation{%
  \institution{UCL Interaction Centre}
  \city{London}
  \country{United Kingdom}
}

\renewcommand{\shortauthors}{Xu et al.}

\begin{abstract}
External Human-Machine Interfaces (eHMIs) have been proposed to enhance communication between automated vehicles (AVs) and pedestrians, with growing interest in multi-modal designs such as audio-visual eHMIs. Just as poor lighting can impair visual cues, a loud background noise may mask the auditory stimuli. However, its effects within these systems have not been examined, and little is known about how pedestrians --- particularly Deaf and Hard-of-Hearing (DHH) people --- perceive different types of auditory stimuli. We conducted a virtual reality study (Hearing N=25, DHH N=11) to examine the effects of background noise (quiet and loud) on auditory stimuli (baseline, bell, speech) within an audio-visual eHMI. Results revealed that: (1) Crossing experiences of DHH pedestrians significantly differ from Hearing pedestrians. (2) Loud background noise adversely affects pedestrians' crossing experiences. (3) Providing an additional auditory eHMI (bell/speech) improves crossing experiences. We outlined four practical implications for future eHMI design and research. 
\end{abstract}




\maketitle

\section{Introduction} 
When crossing in front of vehicles, pedestrians rely on both implicit (e.g., speed, distance, and deceleration) \citep{8241847, SCHMIDT2009300} and explicit communication (e.g., driver eye contact, hand gestures, or headlight flashes) \citep{SUCHA201741}. 
In automated vehicles (AVs), the absence of a human driver removes these conventional channels. This creates two challenges:
(1) The loss of explicit driver signals may increase uncertainty and ambiguity in interactions; (2) people are often poor at judging implicit communication cues, such as speed, stopping distance, and time-to-arrival~\citep{Lee2019,SUN201597, PETZOLDT2014127}. To address these challenges, researchers have proposed external human-machine interfaces (eHMIs) to facilitate communication between AVs and vulnerable road users, improving safety, subjective crossing experiences, and behaviour \cite{doi:10.1177/0018720819836343, 10.1007/978-3-319-41682-3_41, scalability_Mark, multi_modal_dey}. 

Yet, eHMI research has rarely involved disabled people, excluding an estimated 430 million (over 5\% of the world’s population) deaf and hard-of-hearing (DHH) people in the world and potentially 700 million (i.e., 10\% of the whole population) DHH people in 2050\footnote{\href{https://www.who.int/news-room/fact-sheets/detail/deafness-and-hearing-loss}{WHO: Deafness and Hearing Loss}; accessed 14.08.2025}. A report from the UK has found that disability of pedestrians is a critical contributory factor to fatal or serious collisions with pedestrians\footnote{\href{https://www.gov.uk/government/statistics/reported-road-casualties-great-britain-pedestrian-factsheet-2022}{Reported road casualties in Great Britain: pedestrian factsheet, 2022}; accessed 14.08.2024}, eHMI design should consider the needs of disabled people to ensure the proposed eHMI is accessible to them and create equality in transport.

One proposed approach is multi-modal eHMI~\citep{Asha_wheelchair,mark_include_impairment, Mark_Vision, 10.1145/3546717}. For instance, it could be an audio-visual eHMI with visual eHMI serving as the foundation~\citep{10555703, multi_modal_dey,Xu_CHI_26_towards} while using auditory eHMI to enhance perceived safety \citep{sym13040687}. The choice of the auditory eHMI could also benefit low vision and blind people \citep{Mark_Vision, GoogleWaymo2022} or those who experience situational impairments (e.g., being distracted by secondary activities, occluded view, etc.)~\citep{10.1145/3349263.3351523}. However, the value of audio-visual eHMI, especially auditory stimuli among DHH people, remains unclear. Environmental factors such as background noise further complicates the issue. Ambient background noise could interfere with pedestrians' ability to detect and localise vehicles, increasing risks of injuries \citep{Ambient_Noise_Mask}. Loud background noise could become excessive and obscure auditory signals intended for pedestrians \citep{Mahadevan_audio}. Despite this, only 29\% of eHMI studies have included environmental noises (i.e., natural sounds and human-produced sounds), typically to enhance the simulation realism rather than to examine their interaction with auditory stimuli~\citep{tran20201review}. The investigation into how practical and useful auditory stimuli are under different noise levels for both hearing and DHH people is overlooked, which is the main gap we address in this paper. In the meantime, we aim to identify accessibility barriers that auditory eHMIs may introduce and ensure that the use of auditory cues does not inadvertently disadvantage DHH people.

\medskip

\noindent\fcolorbox{orange}{orange!30}{\textbf{Contribution Statement}~\cite{Wobbrock.2016}}

\smallskip

The main contributions of the paper include: (1) the first empirical virtual reality (VR) simulation evaluation of the effects of background noise (\textit{Quiet}, \textit{Loud}) on auditory stimuli (\textit{Baseline}, \textit{Bell}, \textit{Speech}) on crossing experiences (trust, acceptance, perceived safety, mental load) and behaviour (eye gaze, step into the road time, early step into the road count) between hearing participants (N=25) and DHH participants (N=11); (2) four practical implications that pave the way for future eHMI design and research.

\section{Related Work}
\subsection{DHH People and Road Crossing}
Both hearing and eyesight are important for acting and responding adequately in traffic situations~\citep{deafblind}. DHH pedestrians would face similar visual challenges as Hearing pedestrians when crossing the road (i.e., obstructed views, low lighting, adverse weather conditions, dazzling)~\citep{wearther_lighting, crossing_visual}. However, they face more difficulties concerning sound direction and distance judgments due to hearing loss, which is essential to judge the location of potential threats or obstacles~\citep{kolarik2015auditory,hearing_aids_can_fail}. A survey by ~\citet{GUR2021100994} suggested that almost half of DHH teenagers were involved in traffic accidents as a pedestrian, 2-3 times higher than hearing teenagers. A field study in an urban environment with manually driven vehicles by \citet{street_crossing_disability} showed that DHH pedestrians often experienced heightened apprehension when initiating a crossing and exercised greater caution toward approaching vehicles to ensure safety. \citet{hearing_aid_location} suggest that pedestrians with moderate deafness are at a higher risk of being injured by a vehicle because they have difficulty in identifying the sound direction. Overall, the lack of access to auditory information has reduced feelings of safety among DHH pedestrians~\citep{deaf_road_safety}. Over time, this could constitute a negative and fatiguing experience, discourage active travel activities like walking~\citep{Active_travel}, resulting in reduced physical activity levels~\citep{Carlin2016} and broadening inequities in mobility~\citep{inequality_mobility_walking}. 

According to the World Health Organization~\citep{WHO2021HearingReport}, hearing aids and hearing implants are commonly used hearing technologies to help DHH people with hearing aids mainly intended to help people with mild to moderate hearing loss\footnote{\url{https://www.nidcd.nih.gov/health/hearing aids}; accessed 14.08.2025} and hearing implants to help provide a sense of sound to people with severe to profound deafness\footnote{\url{https://www.nidcd.nih.gov/health/cochlear-implants}; accessed 14.08.2025}. However, these technologies come with their own issues. Hearing aids have consistently failed to improve sound localisation and could even impair it~\citep{hearing_aids_can_fail}. Studies suggest that only about one in five people who would benefit from a hearing aid use one, potentially due to comfort level and perceived benefits not meeting the expectation~\citep{McCormack01052013}. As for hearing implants, they process sounds electronically and transmit electrical stimulation to the cochlea of the individual with hearing impairment, restoring some sensation of auditory perception to help them understand sounds or speech~\citep{Cochlear_implants}. However, data from the UK showed that only 1.3\% of individuals with severe or profound deafness use hearing implants \footnote{Hearing loss statistics in the UK: 
\url{https://www.hearinglink.org/your-hearing/about-hearing/facts-about-deafness-hearing-loss/}; accessed 14.08.2025}. In summary, neither is the technology flawless, nor does it completely compensate for the lack of auditory cues for DHH people. Therefore, we employ the audio-visual eHMI so that visual information is still available for people when they may miss essential auditory information. 

\subsection{Effect of Background Noise and Road Crossing}

Background (or ambient) noise has an impact on drinking behaviour \citep{McElrea1992FastMusic}, reading behaviour \citep{KALLINEN2002537}, office-related tasks \citep{BanburyBerry1998Disruption}, and perception of time in gyms \citep{Musical_tempo}. Similar results were found in mobile interaction, where outdoor urban background noise led to more errors in the visual search task (i.e., finding an icon) and longer time for the text entry task when compared to indoor urban background noise \citep{Noise_Type_Mobile}. Background noise could also affect pedestrian activities such as walking along the road or crossing the road. \citet{traffic_noise_human} found that loud traffic noise levels and higher densities of traffic reduced pedestrians' awareness of objects placed along their route, to walk faster, and to engage more in a straight-ahead gaze fixation. \citet{TAPIRO2018219} found urban background noise plays a negative role in pedestrian crossing behaviour, although visual distraction affected more. In this work, we were interested in how background noise in a typical urban environment (e.g., construction, social activities \citep{Vianna2015Noise}) would impact interaction between AV and pedestrians.

\begin{figure*}
 \centering
 \includegraphics[width=1\textwidth]{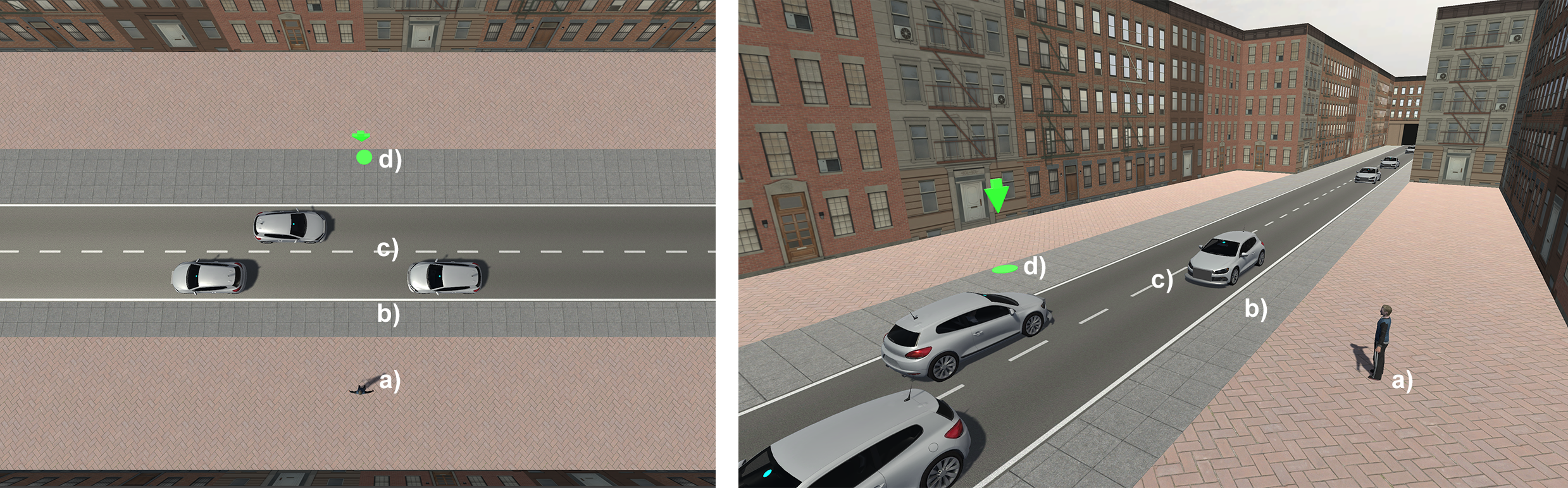}
 \caption{Setup of the virtual simulation environment: The left figure shows a top-down view, while the right figure shows a leftward view. a) The starting position of the participant, which is 4.7 m from the road. b) The 2 m wide grey pavement area, which has a trigger to inform one of the AVs to yield for the participant. c) The 6.8 m wide two-lane road that the participant needed to cross. d) The waypoint (i.e., green surface with downward arrow indicated), which is 2.5 m away from the road. Please note that the human model of the player was only a visual aid to help understand the location; the human model was disabled in the environment, and participants do not have their own avatars.}
  \label{fig:environment}
  \Description{Two images show a city street with parked cars and a pedestrian, labeled a) starting point, b) 2m grey pavement, c) two-lane road, d) waypoint, viewed from above (left) and from street level (right) with arrows indicating directions.}
\end{figure*}

\subsection{Auditory eHMIs}\label{Auditory eHMI Literature}
Auditory eHMIs are typically presented through speech (i.e., verbal messages) \citep{Mahadevan_audio, DEB2018135, Hudson_verbal} or non-speech (e.g., jingles, humming, bell) stimuli \citep{Bell_sound,multi_modal_dey, Florentine}. Through a video-based study, \citet{multi_modal_dey} found that different participants had completely different associations and mental models for using non-speech stimuli (i.e., bell and drone-like humming). While many people perceived the bell as a calm, inviting, and friendly signal, others experienced it as urgent and linked it with a warning. Similarly, although some people felt that a drone (i.e., humming) sound for the vehicle's engine was a natural and fitting choice, others considered it unpleasant and burdensome. \citet{10.1145/3568162.3576979} investigated non-speech stimuli (e.g., humming, bell, jingle) in the wild and found that the humming sound was good for showing the presence but insufficient for other purposes. In contrast, the repeated bell was found to be powerful in indicating the vehicle's stopping. 

On the other hand, speech might be more reliable than non-speech auditory sounds. \citet{DEB2018135} investigated multiple auditory stimuli in a VR-simulated crossing study, finding that the speech was the most favoured auditory stimuli compared to horn, music, and no sound. This is supported by \citet{Hudson_verbal}, where they found speech was preferred over music. Nevertheless, none of these studies have included DHH people, nor explored how auditory stimuli would perform under different noise levels. To the best of our knowledge, this is the first study of its kind. We followed \citep{asha2022towards} to employ a speech eHMI and a bell eHMI (non-speech) in our study as the starting point of this type of research. 

\subsection{Multi-modal eHMIs: Accessible eHMI Solutions}\label{multi-modal_related_work}
Only limited eHMI research has been conducted with disabled people. Some works focus on co-designing personalised solutions based on pedestrians' own devices \citet{Asha_wheelchair}. Others explored the potential of making on-vehicle eHMI accessible to wider access for disabled people. \citet{Mark_Vision} explored the choice of auditory eHMIs with low vision and blind people through a workshop study and found that the speech auditory eHMI was best received. Their follow-up study through VR simulation suggests that having more content in the speech message could reduce the mental load. \citet{10.1145/3546717} compared eHMI concepts (baseline, visual-only, auditory-only, multi-modal---both audio-visual) through an online video-based survey study with participants with intellectual disabilities and participants without intellectual disabilities; they found that multi-modal eHMIs positively affect quality and inclusion \citet{10.1145/3546717}. 

Multi-modal eHMI could also benefit wider pedestrian group when compared to uni-modal eHMI. \citet{sym13040687} conducted a VR study to evaluate 12 eHMI concepts, in combinations of visual (smile/arrow), audio (human voice/warning sound), and vehicle movement style (the approaching speed decreases gradually/remains unchanged), and concluded that multi-modal eHMIs resulted in more satisfactory interaction and improved safety compared to the unimodal eHMI. \citet{He_VR} conducted a VR study with 12 participants and found that audio-visual modality (symbol and anthropomorphic voice) was more appealing than the eHMI with a single modality (visual or auditory). Results from the Wizard-of-Oz study by \citep{10.1007/978-3-030-78645-8_27} suggested that a combination of audio-visual modality is most effective in understanding information. 

The need to implement multi-modal eHMIs has been (1) a common agreement among studies involved with disabled people \citep{Asha_wheelchair,mark_include_impairment, Mark_Vision, 10.1145/3546717}, (2) suggested by a recent review paper on accessibility of eHMI concepts \citep{10555703}, and (3) identified as key opportunity to address gaps for disabled people \citep{10309535}. A core reason is that each modality has specific trade-offs~\cite{smartphone}. Multi-modal interfaces or feedback designs have been commonly used beyond the field of eHMI research to make the interaction accessible for disabled people \citep{Argyropoulos2008,10.1145/638249.638260, Covarrubias01072014}. Therefore, in this work, we decided to make the evaluation more practical by employing audio-visual eHMI, i.e., having a constant visual eHMI in addition to the auditory eHMIs (i.e., Bell and Speech; throughout the paper, we considered the baseline condition as a natural auditory stimuli from the running vehicle) that we wanted to explore. Details of the visual eHMI discussion can be found in Section \ref{Visual eHMI Discussion}.

\section{Virtual Simulation Environment} 
We used Unity v2022.3.44 to develop the virtual simulation environment, featuring a straight, two-way, two-lane urban road (see \autoref{fig:environment}). In the UK, where the study was conducted, traffic law dictates drivers to yield at zebra crossings and at red traffic signals, which would predetermine pedestrian expectations of the vehicle’s behaviour based on right of way. To focus on the impact of the eHMI, we therefore used a mid-block location without markings, where pedestrians may legally cross but are advised to do so with caution. This type of setting is widely used in VR studies of pedestrian–vehicle interaction~\citep{ackermans2020effects, tran20201review}, as it requires participants to attend to approaching vehicles and their communication rather than rely on traffic control infrastructure.

The maximum speed was set at 50 km/h, while the simulated AV drove in a range of 40 km/h to 50 km/h. We set 13 AVs to drive in the first lane (i.e., the one closer to the participant). When one AV exit through the tunnel on the right-hand side, a new AV would be initialised to enter the first lane through the tunnel on the left. The time gap between each AV was around 2.4 seconds to 2.8 seconds. In the second lane, 2 AVs drive towards the right-hand side tunnel when launching the environment; these 2 vehicles would disappear after entering the tunnel and leave the second lane empty. This setting for the second lane was to raise awareness among the participants of vehicles potentially approaching from the left-hand side. Simultaneously, we did not want any vehicles in the second lane to affect the crossing decision and behaviour of the participants. 

Participants started at the position shown in (\autoref{fig:environment}a), walked and waited at the grey pavement (see \autoref{fig:environment}b) until they felt it was safe for them to cross the road (see \autoref{fig:environment}c) and reached the green waypoint (see \autoref{fig:environment}d). Each participant crossed twice for each condition.

\subsection{AV}
\subsubsection{Appearance}
\begin{figure}[b]
 \centering
 \includegraphics{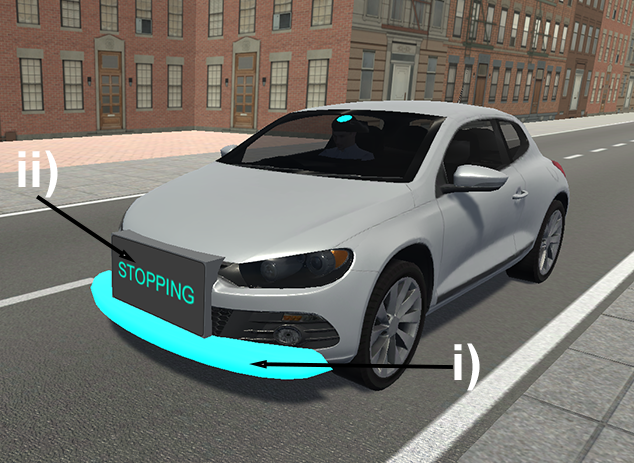}
 \caption{The appearance of the AV used in our study, where a human avatar sits in the driver's seat but does not engage with the vehicle. Active mode of i) the Light Strip and ii) the Display when the AV was fully stopped.}
 \label{fig:AV}
 \Description{A car on a city street with labelled arrows pointing to its front bumper i) Light Strip and ii) Display.}
\end{figure}

\autoref{fig:AV} shows the AVs used in the user study. In line with prior work~\cite{FAAS2020171, Mahadevan_audio, scalability_Mark}, (1) a round cyan light positioned at the top centre of the windshield indicates that the AV is driving autonomously, (2) a light strip located at the bottom of the bumper region displays the light design, and (3) a display located at the grill region of the bumper to display the text/symbol. Unless the AV yielded for participants, neither the light strip nor the display would show any additional information. The visual eHMI was positioned on the front grille of the vehicle, which (1) aligns with current pedestrian expectations and experience, as they typically look towards the location of the driver’s head or vehicle movement \citep{info11010013,10.1145/3342197.3344523}, and (2) follows standard practice of the eHMI research \citep{multi_modal_dey, scalability_Mark}. 

\subsection{Parameters Testing}
We conducted an iterative pilot testing with eight testers (4M; 4F; including 1 DHH tester with severe hearing loss of both ears) to ensure that (1) the visual eHMI design is reasonable and (2) both the choice and volume set for the two urban background noise conditions are reasonable. In addition, they were also satisfied with the other parameters set in the environment, e.g., choice of words and audio, message repetition, interval of the message, tire-road surface sound, and Acoustic Vehicle Alerting System (AVAS). We did not recruit these testers for the formal user study. In our formal user study, none of the participants argued that the background sound and vehicle sound were not realistic.

\subsubsection{Vehicle Sound Implementation}\label{vehicle_sound}
\textbf{AVAS}: We embedded the AVAS sound from BMW into all vehicles for their slowdown process, activated when their speed is from 20 km/h to 0 km/h. The original sound showed a spectrum and frequency shift. When fully stopped, we loop the last second of the sound where they were in a similar frequency range. The average volume of the AVAS was about 64 dB (larger than the minimum requirement of 56 dB). Our implementations followed the regulation \citep{UNECE_WP29_194} and a similar implementation by \citet{multi_modal_dey}; refer to drone auditory eHMI.


\textbf{Tire-Surface}: To improve the realism of the study~\citep{tran20201review}, we also implemented a tire-surface sound system attached to each AV. We ensured the sound\footnote{Tire rolling sound: \url{https://drive.google.com/file/d/1SsnY6dP10cM6LmpIZsTqE5y-3L_Q_taM/view?usp=sharing}; accessed: 09.04.2025} had a volume of 62 dB when driving at 50 km/h. To simulate a realistic slowdown tire-surface effect, we took approximate values found in \citep{Pallas_tire, Iversen_tire}, applied a feed-out effect when the speed reached around 20 km/h (50 dB), and another feed-out effect so the sound reached 0 dB when full stopped.

\subsection{Visual eHMI}\label{Visual eHMI Discussion}
As a visual eHMI represents the standard of the field, we included a fixed visual eHMI candidate across all experimental conditions. We believed multi-modal eHMI should be the priority eHMI representation as (1) its advantage over unimodal eHMI \citep{sym13040687, He_VR,10.1007/978-3-030-78645-8_27} and (2) the potential to be more accessible to disabled pedestrians \citep{Mark_Vision,10.1145/3546717}. Since there is still no explicit agreement on which visual design is the most beneficial and how many visual signals might be suitable or sufficient for the other road users~\cite{multi_modal_dey}, we employed a combination of Abstract Light \cite{Bumper_Light} and Text \cite{text_light} as our visual eHMI concept, this design was widely accepted by participants in prior work \citep{Xu_CHI_26_towards}.

The visual eHMI remained inactive during standard driving. When a yielding command is triggered, the light strip would pulse in Cyan with its pulse frequency going from fast (pulsates between on and off at a rate of 1 Hz) to slow (pulsates between on and off at a rate of 0.5 Hz) to indicate a speed change. Meanwhile, during this slowdown, the display would show "STOPPING" to convey that the vehicle is slowing down. When the AV fully stopped, the light strip would stop pulsing and remain static; meanwhile, the text displayed on the display would change to "STOPPED". Both words were presented in bold cyan letters.

\subsection{Background Noise}
We implemented two typical urban background noise environments: quiet and loud.

\textit{Quiet:} This condition represents a quiet part of the city with relatively light social activities around the environment \citep{Vianna2015Noise}. We used two audio clips \footnote{Urban Ambient Noise 1: \url{https://sound-effects.bbcrewind.co.uk/search?q=07027128} and Urban Ambient Noise 2: \url{https://sound-effects.bbcrewind.co.uk/search?q=07056053}; accessed 29.04.2025} from BBC Sound Effects, with each background noise audio source located near each side of the building. These files form a mixture of overlapping human conversation (e.g., conversational speech, laughter, calls) and incidental background clutter. Standing at the starting point, the perceived sound volume was approximately an average of 58 dB with maximum volume reached around 64 dB, which is roughly the standard city noise sound \citep{10.1145/3409120.3410646,CPUC2011CresseyGallo} and is the volume implemented by prior work in eHMI literature \citep{8957075}. 

\textit{Loud:} This condition represents a noisy part of the city with dense social activities and a construction site \citep{Vianna2015Noise}. We used three audio clips \footnote{Urban Ambient Noise 1: \url{https://sound-effects.bbcrewind.co.uk/search?q=07056049}, Urban Ambient Noise 2: \url{https://sound-effects.bbcrewind.co.uk/search?q=07001118}, and Construction: \url{https://freesound.org/people/klankbeeld/sounds/348624/}--we amplified this sound to be around 80 dB; accessed: 09.04.2025} from BBC Sound Effects and Free Sound organisation. In line with the \textit{Quiet} condition, each background noise audio source is located at one side of the building. The construction file was located 1 meter next to the background noise audio source at the same end of the starting point. These files combine construction-site activity (i.e., machinery, hammering, and high-frequency industrial sounds) and a busy market environment (i.e., dense human chatter, crowd movement, and general ambient bustle). Standing at the starting point of the environment, the participant would perceive an average sound volume of 70 dB, with the maximum volume reach 79 dB \citep{10.1145/3409120.3410646}. A real world city environment may consist of even louder sound or sudden unexpected sound \citep{Reddy2025NoiseHazard}, however, we did not include this and control this to be the volume as we described as (1) exposing loudness of over 85 dB could cause hearing loss \citep{SliwinskaKowalska2012} which is not ethical, and (2) testing sudden unexpected sound is beyond the focus of this study.

To mitigate participant familiarity with the background noise and introduce natural variation across trials, we randomised the playback onset of each audio source. Specifically, the system initiated playback at a random point within the first 0 -- 30 seconds of the recording.

\subsection{Auditory Stimuli}
As discussed in the Section \ref{Auditory eHMI Literature}, we selected \textit{Speech} and \textit{Bell} (no-speech) as our Auditory eHMIs.

\textit{Speech:} \citet{status_zhang, FAAS2020171}, ISO technical report \citep{iso23049}, and Volvo\footnote{\href{https://www.media.volvocars.com/global/en-gb/media/pressreleases/237019/volvo-360c-concept-calls-for-universal-safety-standard-for-autonomous-car-communication1\#}{Volvo: Volvo 360c Concept}} suggested that AVs should not give cross advice as it creates liability and legibility issues. Therefore, we used intention-based messages such as ``I'm stopping'' and ``I'm stopped'' for the slowdown and fully stopped state, respectively. These messages were generated by an online text-to-audio website\footnote{\url{ttps://ttsmp3.com/} with US English / Kendra accent; accessed: 14.05.2025}. When a yielding command was activated in the AV, it would first wait for 1.5 seconds and then play ``I’m stopping'' (0.75 seconds long) with a 0.5-second interval. The choice of a 1.5-second delay is to allow the AV to get closer to the pedestrian. This message would play three times during the slowdown process. When the AV fully stopped, the AV would immediately play the message ``I’m stopped'' (0.7 seconds long) with a 1.1-second interval. The AV would keep repeating this verbal message with a proposed interval until the participant passed the first lane of the road.

\textbf{Bell:} We employed a Bell sound downloaded from BBC Sound Effects\footnote{Bell: \url{https://sound-effects.bbcrewind.co.uk/search?q=07066164}}, which is a similar Bell sound as in \citep{multi_modal_dey}. Following the implementation from \citep{multi_modal_dey}, the Bell sound (1.1 seconds long) used for stopping and full stop was the same. When a yielding command was activated, the AV would wait for 1.5 seconds and then play the Bell sound with a 0.3-second interval. The AV would play the Bell sound three times during the slowdown process. When the AV fully stopped, the AV would immediately play the Bell with a 1-second interval to indicate the difference between stopping and being stopped. Similar to the speech auditory design, the AV would repeat this bell sound with a proposed interval until the participant passed the first lane of the road. 

All files have an average volume of around 65 dB, with the maximum volume of around 74 dB. We tested it by placing the audio source 2 meters away from the main camera in Unity, and placed a volume measurement device in the middle of the headset.


\section{Virtual Reality User Study}
This study was guided by the following research questions (RQs):

\begin{quote}

\textbf{RQ1}: How do the ratings for experience (i.e., trust, acceptance, perceived safety, mental load) and behaviour (i.e., gaze behaviour, step-in road time, early step into the road count) differ between Hearing and DHH participants?

\textbf{RQ2}: What impact does the Background Noise have on pedestrians regarding experience and behaviour? 

\textbf{RQ3}: What impact do the Auditory Stimuli of the Audio-Visual eHMI have on pedestrians regarding experience and behaviour? 
\end{quote}

\subsection{Study Design and Outcome Measures}\label{Questionnaire}
We employed a 2 $\times$ 3 within-subjects design with two within-subjects factors (1) Background Noise (Quiet and Loud) and (2) Auditory Stimuli (Baseline, Bell, Speech). The order of Background Noise $\times$ Auditory Stimuli was counterbalanced in the study.

\textit{Subjective}: Crossing experience was measured via questionnaires after each condition. 
\begin{itemize}
  \item We used a 21-point single question regarding the mental workload from the NASA-TLX~\cite{HART1988139} questionnaire to measure mental workload (the lower the number, the smaller the workload).
  \item We employed the 5-point Likert scale Trust in Automation questionnaire~\cite{trust_korber} to measure the trust in AV (Understandability and Trust) through the 4-item Understandability and 2-item Trust subscales (the higher the number, the better).
  \item Perceived Safety was measured by a 7-point Likert scale that ranged from -3 (anxious/agitated/unsafe/timid) to +3 (relaxed/calm/safe/confident) ~\cite{safety_Faas}.
  \item We employed the van der Laan acceptance scale with the subscales `usefulness' and `satisfying' to measure Acceptance~\cite{VANDERLAAN19971}. 
\end{itemize}

\textit{Objective}: Crossing behaviour was recorded by the developed program. For eye gaze, we counted the fixations when the AV was 20 meters away from the participant's crossing point till the participant stepped into the road. We choose 20 meters as the starting point for calculating the results in this paper because early research suggests that pedestrians start to look more at the vehicle than at the road ahead from this range \citep{10.1145/3342197.3344523} and it provides higher eye gaze accuracy. Light Strip, Display, and the Whole Vehicle are the areas of interest we are keen to explore as they allow us to understand if gaze behaviour of participants would change under different sound and noise conditions when interacting with the eHMI (e.g., would they pay more attention to the vehicle in general or the active eHMI component due to its perceptual salience) \citep{TAPIRO2018219}.

\begin{itemize}
    \item Step Into the Road Time: The time taken by the participant to start crossing the road from the moment an AV starts to slow down on the nearest lane.
    \item Early Step Into the Road Count: The number of times the participant stepped onto the road before the AV fully stopped.
    \item Eye Gaze 
    \begin{itemize}
     \item \textbf{Light Strip Duration}: Fixation on the Light Strip measured in second.
     \item \textbf{Display Duration}: Fixation on the Display measured in second.
     \item \textbf{Whole Vehicle Duration}: Fixation on the combination of Light Strip, Display, and other parts of Vehicle, measured in second.
     \item \textbf{Active Visual eHMI Duration}: duration on active visual eHMI for conditions that employ Visual eHMI, i.e., Light Strip for \textit{Abstract Light}, combination of Light Strip and Display for both \textit{Abstract Light + Text} and \textit{Abstract Light + Symbol} conditions.
     \item \textbf{Active Visual eHMI Percentage}: (Active Visual eHMI Duration $/$ the Whole Vehicle Duration) $\times$ 100\%.
   \end{itemize}
\end{itemize}

At the end of the study, participants rated their perception of the Necessity and Reasonability of (1) Bell and (2) Speech concepts in a 7-point (1=Totally Disagree to 7=Totally Agree)~\cite{scalability_Mark} and then ranked the Audio eHMIs (i.e., Baseline, Bell, Speech), the lower the number, the better. This was followed by a semi-structured interview. We first asked the participants questions about the Baseline--- ``Were you able to hear and understand?'' and ``Do you think the background sound impacts how you perceive it?'' Then the interview moved to the proposed auditory eHMI design---``Could you hear it when the vehicle (1) slowed down and (2) fully stopped?'', ``Overall, what do you think about it?'', ``Anything you liked/disliked about it?'', and ``Do you think the background sound impacts how you perceive it?''. The final stage of the interview followed two questions ``Can you share some examples of sound that affect you in a road crossing or street walking activity?'' and ``Would you think you can cross without the auditory eHMI?'' In the end, we offered an open question to ask if participants had anything to add.

\subsection{Apparatus and Setup}
A Varjo XR-4 focal edition was used as the VR headset, which offers a 90 Hz refresh rate with 3840 $\times$ 3744 resolution and a 120° $\times$ 105° field of view. We enabled the built-in 200 Hz eye tracker during the study, which provides and records eye gaze visualisation and eye measurements such as pupil iris diameter, openness, and interpupillary distance~\footnote{\href{https://developer.varjo.com/docs/unity-xr-sdk/eye-tracking-with-varjo-xr-plugin}{Varjo Developer Eye Tracking}}. Cyberith Virtualizer Elite 2 was used as our walking solution for locomotion. Varjo XR-4 and Cyberith Virtualizer Elite 2 were connected to a high-end PC with an i9 CPU, 64 GB RAM, and a GeForce RTX 4090 Graphics card to provide the best immersive experience. The sound experience was provided directly through the VR device's built-in speakers, as wearing additional headphones or earphones would clash with hearing aids or implants and may result in an uncomfortable experience. The study was conducted in an indoor, well-illuminated, quiet laboratory room that could not be seen from outside.

\subsection{Procedure}
The study started with a brief introduction from the experimenter. Subsequently, the participants needed to sign the consent form and complete a demographic questionnaire. Participants were given two trials to get familiar with the Virtualizer Elite 2 device and the Varjo XR-4. Once the participants were ready, they needed to complete the formal study conditions (we counterbalanced the order across participants), with their task being to cross the street twice. Eye calibration was made/checked at the beginning of each condition to ensure accuracy. After each condition, participants had to answer the required questionnaires (See Section \ref{Questionnaire}). At the end of the study, they participated in a semi-structured interview, which lasted about 10-15 minutes. The study lasted about 60 minutes for English users and 90 -- 120 minutes for sign language users (longer time for questionnaire and interview). Participants were under the observation and supervision of an experimenter. A BSL interpreter was always present to assist with the study for BSL users.

\begin{table*}
\caption{Demographic Information of the DHH Participants.}
\begin{tabular}{p{0.05\linewidth} p{0.05\linewidth} p{0.075\linewidth} p{0.2\linewidth} p{0.1\linewidth} p{0.23\linewidth} p{0.13\linewidth}}
\toprule
    ID & Age & Gender & Left \& Right Ear \newline Hearing Loss Level & Identity & Preferred Communication & Using Hearing Technology?\\
    \midrule
    P25 & 33 & Female & Profound; Profound   & Deaf & BSL & Yes\\
   P26 & 71 &	Male & Severe; Severe   & HoH & English &	Yes	\\
   P27	& 41 & Male &	Profound; Profound & Deaf & BSL &	Yes \\
    P28 &	60	& Female & 	Profound; Severe & HoH & English &	Yes\\
    P29 &	78	& Male &	Severe; Severe & HoH & English &	Yes\\
    P30 &	21	& Male	& Profound; Profound & deaf & English & Yes\\
    P31 &	69	& Female &	Moderate; Severe  & HoH & English &	Yes\\
    P32 &	71	& Female & 	Moderate; Severe & deaf & English &	Yes\\
    P33 &	60	& Male	& Profound; Profound & deaf & BSL &	Yes\\
    P34	& 34 & Female &	Severe; Severe & deaf & English&	Yes\\
    P36 &	 53	& Female	& Severe; Profound & deaf &English  & Yes\\ 

    \bottomrule
    \end{tabular}
  \label{table:VR_DHH}
  \Description{Table shows demographic of deaf and hard of hearing participants, with these columns show data from left to right to be: ID, Age, Gender, Left & Right Ear Hearing Loss Level, Identity, Preferred Communication, and whether or not Using Hearing Technology?}
\end{table*}

\subsection{Participants}
Participants were recruited via physical posters, social media platforms, charities, and informal referrals through word-of-mouth. In total, we recruited 36 participants to the user study, with 25 participants (12 male, 13 female; Mean age $=26.4, SD=3.96$, range 21 to 35) who self-identified as no hearing loss people, including 1 participant who had one ear for Mild hearing loss level and 1 participant who had Moderate hearing loss for one ear; and 11 participants (5 male, 6 female; Mean age $=53.73, SD=18.84$, range 21 to 78) who identified themselves as DHH people (see \autoref{table:VR_DHH}). We categorised the hearing loss levels, normal (<20 dB), mild (21 - 40 dB), moderate (41 - 70 dB), severe (71 - 95 dB), and profound (>95 dB), based on the recommendations from the National Health Services\footnote{National Health Service: \url{https://www.esht.nhs.uk/service/audiology/diagnosis-and-testing/}; accessed: 08.03.2025}. All DHH participants participated in our study used a pair of hearing aids or implant during the study. In a 5-point Likert scale (1 = Not at all, 5 = Definitely), Hearing participants showed medium interest in AVs ($M=3.2, SD=1$) and had more knowledge about AVs ($M=4.28, SD=0.84$). In contrast, Participants in DHH reported less interest ($M=2.82, SD=1.40$) and knowledge ($M=3.64, SD=0.92$) in AVs. 

We adhere to all institutional safety measures and data protection guidelines throughout the experiment. We also consulted experienced accessibility researchers who specialise in working with DHH participants to review our study set-up and design to ensure it is accessible and friendly to all participants. Informed consent was obtained from all participants and the study was approved by our university's Research Ethics Committee.

\section{Results}
We focused on the main and interaction effects of the three independent variables: \textit{Background Noise} (\textit{Quiet} and \textit{Loud}; within-subjects) and \textit{Auditory Stimuli} (\textit{Baseline}, \textit{Bell}, and \textit{Speech}; within-subjects), and \textit{Hearing Group} (\textit{Hearing} and \textit{DHH}; between-subjects). Shapiro-Wilk test suggested that our data were not normally distributed. We therefore used \textit{nparLD}, which can handle small samples and unequal group sizes \cite{JSSv050i12}, and has been widely used by similar studies \cite{Mark_Vision,10.1145/3411764.3445597,10.1145/3411764.3445492}. The Modified ANOVA-type statistic are reported for the whole-plot factor (i.e., main effect of \textit{Hearing Group}) as suggested by \citep{Brunner2002}. Otherwise, ANOVA-type statistics are reported by default. We employed Bonferroni correction for all post-hoc tests.

\subsection{Subjective Crossing Experience}

\begin{figure}[t]
  \centering
  \includegraphics[width=\linewidth]{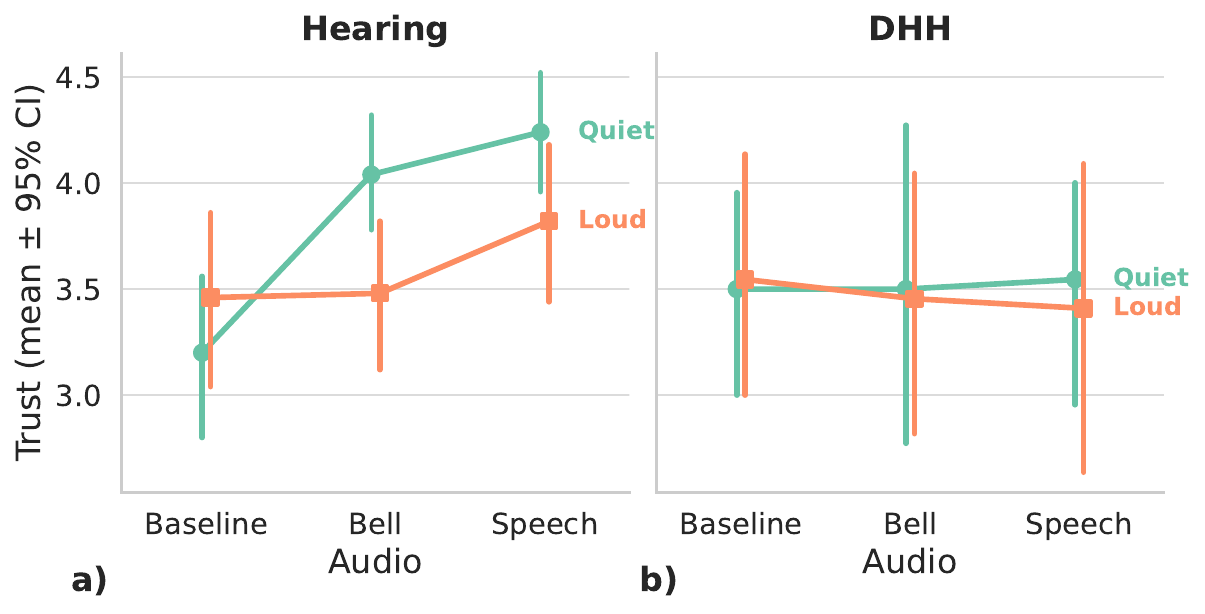}
  \caption{Mean Trust ratings of audio stimuli types under Quiet vs. Loud background noise across a) Hearing participants and b) DHH participants. Data are shown with 95\% confidence intervals. }
  \label{fig:trust}
  \Description{Data figure illustrates how background noise and audio stimuli types differently influence trust rating across groups.}
\end{figure}

\subsubsection{Trust in Automation} 
We could not find any significant difference among the \textbf{Understandability} ratings. 

Regarding \textbf{Trust}, \autoref{fig:trust} illustrates trust ratings for all conditions across \textit{Hearing} and \textit{DHH} participants. The non-parametric variance analysis (NPVA) revealed a significant main effect of \textit{Background Noise} ($F=4.000, df=1, p=.045$) and \textit{Auditory Stimuli} ($F=9.101, df=1.879, p<.001$). Post-hoc analysis for \textit{Background Noise} confirmed that participants gave higher trust ratings in \textit{Quiet} conditions ($M=3.73, SD=0.95$) than \textit{Loud} conditions ($M=3.55, SD=1.00, p=.016$). Post-hoc pairwise comparisons for the main effect of \textit{Auditory Stimuli} showed that participants gave lower rating for \textit{Baseline} ($M=3.39, SD=0.97$) than \textit{Bell} ($M=3.67, SD=0.95, p=.032$) and \textit{Speech} ($M=3.86, SD=0.96, p=.002$). 

The NPVA showed a significant interaction effect of \textit{Hearing Group} $\times$ \textit{Background Noise} ($F=4.047, df=1, p=.044$). Post-hoc results showed that among the \textit{Hearing} group, higher trust ratings were given in \textit{Quiet} conditions ($M=3.83, SD=0.91$) than \textit{Loud} conditions ($M=3.59, SD=0.95, p=.021$). The NPVA also yielded a significant interaction effect of \textit{Auditory Stimuli} $\times$ \textit{Background Noise} ($F=6.309, df=1.837, p=.002$), post-hoc pairwise comparisons showed that (1) among \textit{Bell}, trust was rated higher in \textit{Quiet} conditions ($M=3.86, SD=0.94$) than \textit{Loud} conditions ($M=3.59, SD=0.94, p=.0133$), (2) within \textit{Quiet} conditions, \textit{Baseline} ($M=3.29, SD=0.94$) was rated significantly lower than \textit{Bell} ($M=3.88, SD=0.94, p=.007$) and \textit{Speech} ($M=3.69, SD=0.95, p=.001$). In addition, the NPVA yielded a significant interaction effect of \textit{Hearing Group} $\times$ \textit{Auditory Stimuli} ($F=8.476, df=1.879, p<.001$). Post-hoc results showed that among \textit{Hearing} people, \textit{Baseline} ($M=3.36, SD=0.96$) was rated significantly lower than \textit{Bell} ($M=3.76, SD=0.83, p=.024$) and Speech ($M=3.94, SD=0.91, p=.003$).

\begin{figure}[t]
  \centering
  \includegraphics[width=\linewidth]{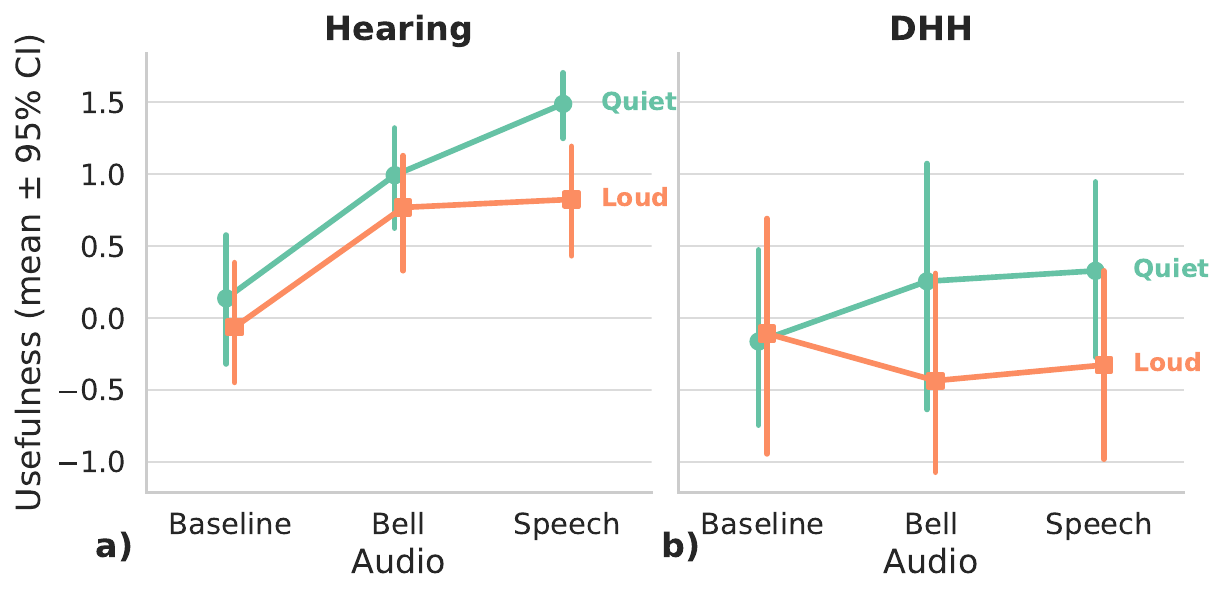}
  \caption{Mean Usefulness ratings of audio stimuli types under \textit{Quiet} vs. \textit{Loud} background noise across a) \textit{Hearing} participants and b) \textit{DHH} participants. Data are shown with 95\% confidence intervals. 
  }
  \label{fig:usefulness}
  \Description{Data figure illustrates how background noise and audio stimuli types differently influence usefulness rating across groups.}
\end{figure}

\subsubsection{Acceptance} 
As for \textbf{Usefulness}, the NPVA revealed a significant main effect of \textit{Background Noise} ($F=10.869, df=1, p<.001$), \textit{Auditory Stimuli} ($F=8.177, df=1.907, p<.001$), and \textit{Hearing Group} ($F=5.042, df=1, p=.040$; Modified ANOVA-type statistic). Post-hoc pairwise comparisons of \textit{Background Noise} showed that participants gave a higher usefulness rating in the \textit{Quiet} conditions ($M=0.65, SD=1.18$) than \textit{Loud} conditions ($M=0.26, SD=1.21,p<.001$). Post-hoc analysis for the \textit{Auditory Stimuli} showed that both \textit{Bell} ($M=0.58, SD=1.19, p=.010$) and \textit{Speech} ($M=0.80, SD=1.09, p<.001$) were better than \textit{Baseline} ($M=-0.02, SD=1.19$). Post-hoc analysis for the \textit{Hearing Group} showed that \textit{Hearing} participants ($M=0.69, SD=1.10$) gave significantly higher ratings than \textit{DHH} participants ($M=-0.08, SD=1.26, p<.001$). 

We found interaction effects of \textit{Hearing Group} $\times$ \textit{Auditory Stimuli} ($F=6.179, df=1.907, p=.002$). Post-hoc results showed that (1) among \textit{Hearing} participants, \textit{Bell} ($M=0.88, SD=0.97, p=.005$) and \textit{Speech} ($M=1.16, SD= 0.88, p<.001$) were rated significantly higher than \textit{Baseline} ($M=0.04, SD=1.14$). (2) Regarding \textit{Bell}, usefulness ratings provided by \textit{Hearing} participants ($M=0.88, SD=0.97$) were significantly higher than \textit{DHH} participants ($M=-0.09, SD=1.40, p=.019$). (3) Regarding \textit{Speech}, usefulness ratings provided by \textit{Hearing} participants ($M=1.16, SD=0.88$) were significantly higher than \textit{DHH} participants ($M=0.00, SD=1.13, p<.001$). \autoref{fig:usefulness} illustrates usefulness ratings for all conditions across \textit{Hearing} and \textit{DHH} participants.

\begin{figure}[t]
  \centering
  \includegraphics[width=\linewidth]{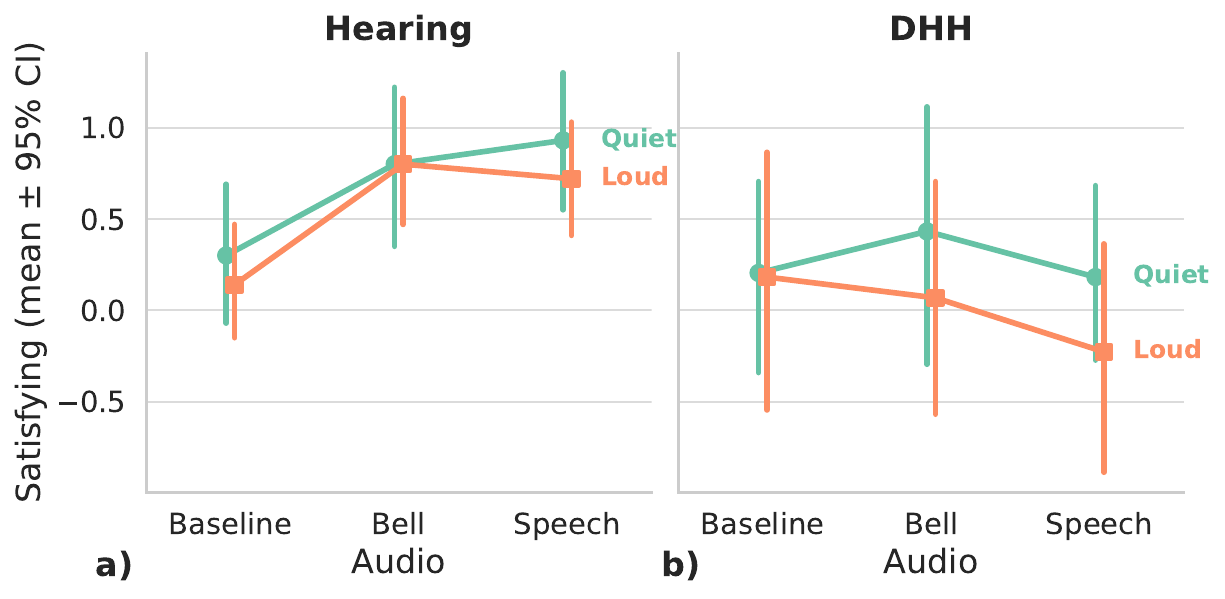}
  \caption{Mean Satisfying ratings of audio stimuli types under Quiet vs. Loud background noise across a) Hearing participants and b) DHH participants. Data are shown with 95\% confidence intervals. }
  \label{fig:satisfying}
  \Description{Data figure illustrates how background noise and audio stimuli types differently influence satisfying rating across groups.}
\end{figure}

Regarding \textbf{Satisfying}, the NPVA revealed no significant main effects. We observed an interaction effect of \textit{Hearing Group} $\times$ \textit{Auditory Stimuli} ($F=4.671, df=1.918, p=.010$). Post-hoc analysis showed that (1) among \textit{Hearing} participants, \textit{Baseline} ($M=0.22, SD=0.88$) was rated significantly lower than \textit{Bell} ($M=0.8, SD=0.99, p=.011$) and \textit{Speech} ($M=0.83, SD=0.89, p=.026$), (2) regarding \textit{Speech}, \textit{DHH} participants ($M=0.05, SD=1.08$) gave a significantly lower rating than Hearing participants ($M=0.83, SD=0.89, p=.003$). Data for each condition across \textit{Hearing} and \textit{DHH} participants can be found in \autoref{fig:satisfying}. 

\subsubsection{Perceived Safety} 
\begin{figure}[t]
  \centering
  \includegraphics[width=\linewidth]{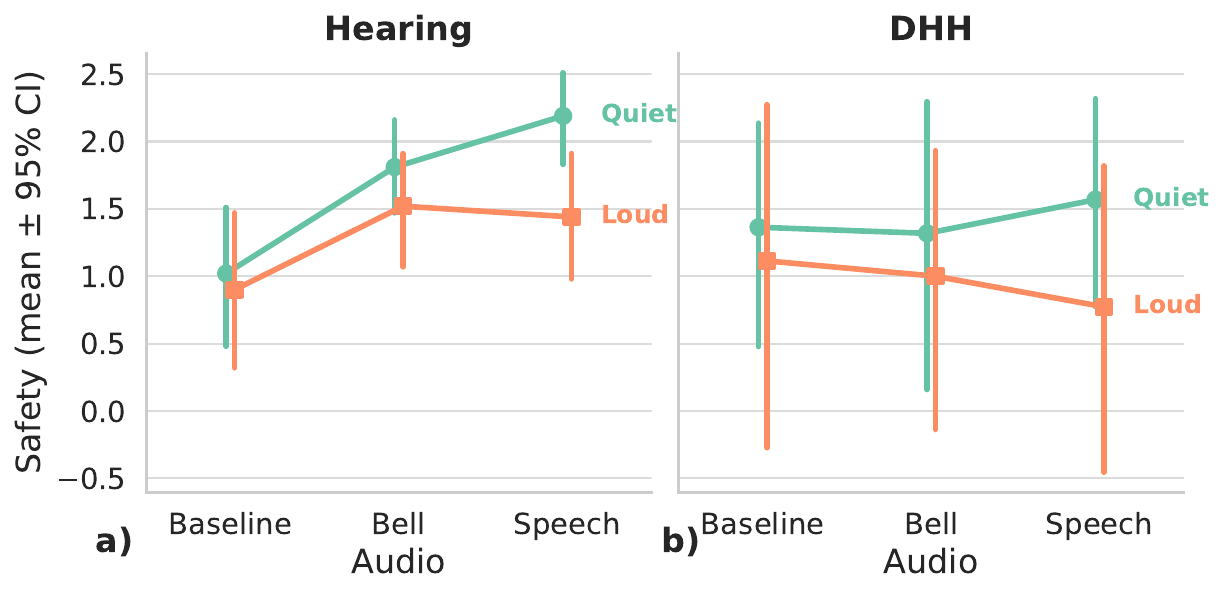}
  \caption{Mean Safety ratings of audio stimuli types under Quiet vs. Loud background noise across a) Hearing participants and b) DHH participants. Data are shown with 95\% confidence intervals.}
  \label{fig:safety}
  \Description{Data figure illustrates how background noise and audio stimuli types differently influence safety across groups.}
\end{figure}

\autoref{fig:safety} illustrates usefulness ratings for all conditions across \textit{Hearing} and \textit{DHH} participants. The NPVA showed a significant main effect of \textit{Background Noise} ($F=12.468, df=1, p<.001$) and \textit{Auditory Stimuli} ($F=4.727, df=1.94, p=.009$). Post-hoc analysis on main effect \textit{Background Noise} confirmed that participants gave a higher rating in \textit{Quiet} conditions ($M=1.59, SD=1.30$) than \textit{Loud} conditions ($M=1.19, SD=1.53, p<.001$). Post-hoc results of the main effect \textit{Auditory Stimuli} showed that ratings for \textit{Baseline} ($M=1.05, SD=1.55$) was significantly lower than \textit{Bell} ($M=1.15, SD=1.34, p=.022$) and \textit{Speech} ($M=1.62, SD=1.35, p=.002$).

We found a significant interaction effect of \textit{Auditory Stimuli} $\times$ \textit{Background Noise} ($F=4.023, df=1.94, p=.018$). Post-hoc pairwise comparisons analysis showed that (a) regarding \textit{Bell}, ratings were higher in \textit{Quiet} conditions ($M=1.66, SD=1.29$) than \textit{Loud} conditions ($M=1.36, SD=1.38, p=.047$) (b) regarding \textit{Speech}, ratings were also higher in \textit{Quiet} conditions ($M=2, SD=1.09$) and \textit{Loud} conditions ($M=1.24, SD=1.49, p<.001$). In addition, post-hoc results also yielded that under \textit{Quiet} conditions, \textit{Baseline} was rated significantly lower than \textit{Speech} ($M=2, SD=1.09, p<.001$). 

We also found a significant interaction effect of \textit{Hearing Group} $\times$ \textit{Auditory Stimuli} ($F=10.791, df=1.94, p<.001$). Post-hoc analysis showed that among \textit{Hearing} group, \textit{Baseline} ($M=0.96, SD=1.40$) was rated significantly lower than \textit{Bell} ($M=1.67, SD=1.05, p=.005$) and \textit{Speech} ($M=1.82, SD=1.10, p=.002$).

\subsubsection{Mental Workload}
\begin{figure}[t]
  \centering
  \includegraphics[width=\linewidth]{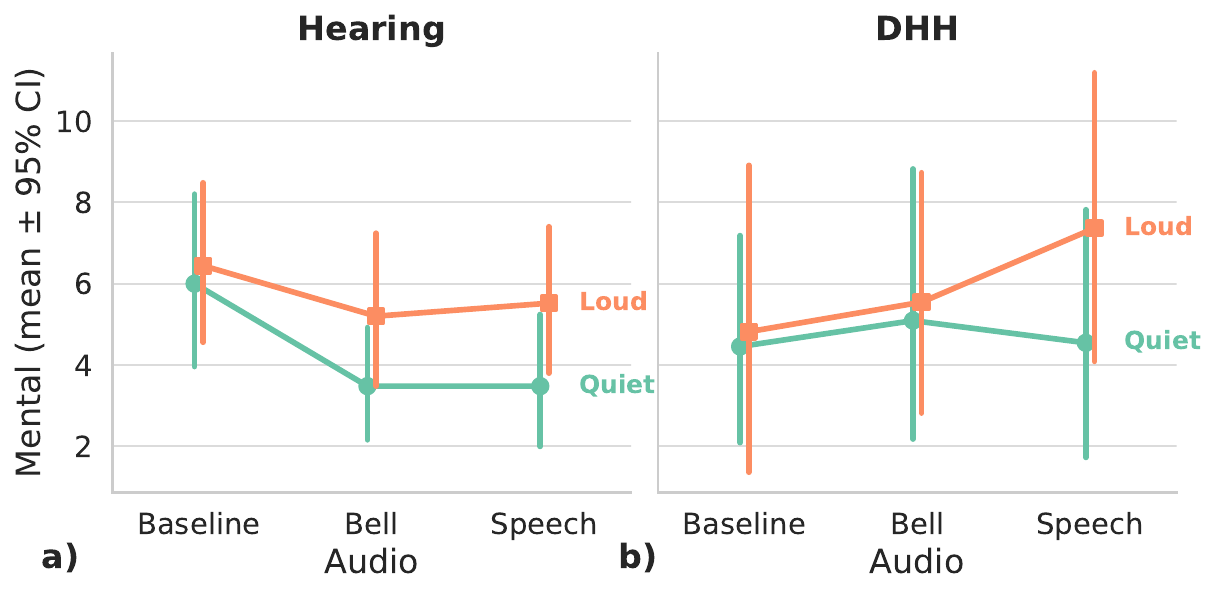}
  \caption{Mean Mental Workload ratings of audio stimuli types under Quiet vs. Loud background noise across a) Hearing participants and b) DHH participants. Data are shown with 95\% confidence intervals. }
  \label{fig:mental}
  \Description{Data figure illustrates how background noise and audio stimuli types differently influence mental workload across groups.}
\end{figure}

The NPVA revealed a significant main effect of \textit{Background Noise} ($F=11.841, df=1, p<.001$) on Mental workload. Post-hoc analysis confirmed that ratings were lower in \textit{Quiet} conditions ($M=4.44, SD=4.75$) than \textit{Loud} conditions ($M=5.78, SD= 5.22, p<.001$). \autoref{fig:mental} shows usefulness ratings for all conditions across \textit{Hearing} and \textit{DHH} participants.

The NPVA showed a significant interaction effect of \textit{Auditory Stimuli} $\times$ \textit{Background Noise} ($F=4.169, df=1.83, p=.018$). Post-hoc pairwise comparisons yielded that (a) for \textit{Bell}, mental workload rating was lower in \textit{Quiet} conditions ($M=3.97, SD=4.46$) than \textit{Loud} ($M=5.31, SD= 4.90, p=.0473$), (b) for \textit{Speech}, mental workload rating was lower in \textit{Quiet} conditions ($M=3.81, SD=4.60$) than \textit{Loud} conditions ($M=6.08, SD=5.13, p<.001$). In addition, post-hoc results also showed that under \textit{Quiet} background environment, mental workload was rated significantly lower for \textit{Speech} ($M=3.81, SD=4.60$) than \textit{Baseline} ($M=5.53, SD=5.12, p=.033$).

The NPVA also yielded a significant interaction effect of \textit{Hearing Group} $\times$ \textit{Auditory Stimuli} ($F=8.844, df=1.89, p<.001$) on mental workload. Post-hoc results showed that among \textit{Hearing} participants, mental workload rating was higher for \textit{Baseline} ($M=6.22, SD=5.30$) than \textit{Bell} ($M=4.34, SD=4.24, p<.001$) and \textit{Speech} ($M=4.5, SD=4.53, p=.007$).

\subsection{Objective Crossing Behaviour} 
\subsubsection{Eye Gaze Behaviour} 
We could not observe any significant effect of \textit{Auditory Stimuli}, \textit{Background Noise}, \textit{Hearing Group} or their interactions among measurements of \textbf{Light Strip Duration}, \textbf{Display Duration}, \textbf{Whole Vehicle Duration}, \textbf{Active Visual eHMI} and \textbf{Active Visual eHMI Percentage}. Contrary to our expectations that \textit{Auditory Stimuli} and \textit{Background Noise} may modulate gaze behaviour, the results indicated that participants' gaze duration for these measurements seems relatively stable across conditions and between Hearing Groups (Hearing and DHH).

\subsubsection{Movement behaviour}
Both data sets below covered 432 trials (36 participants $\times$ 6 conditions $\times$ 2 repetition) of crossings. We did not observe a single crash during the study. 

As for \textbf{Step Into the Road Time}, the NPVA revealed a significant interaction effect of \textit{Hearing Group} $\times$ \textit{Background Noise} ($F=3.872, df=1, p=.049$). However, post-hoc results did not yield any significance. Regarding \textbf{Early Step Into the Road Count}, we also could not observe any significant difference.

\subsection{Necessity and Reasonability for Bell and Speech eHMIs}
Regarding necessity, among Hearing participants, the \textit{Bell} received an average score of 4.96 ($SD=1.51$) and \textit{Speech} received an average score of 5.76 ($SD = 1.36$). As for \textit{DHH} participants, the \textit{Bell} received an average score of 3.91 ($SD=2.17$) and \textit{Speech} received an average score of 4.64 ($SD = 1.63$). Regarding reasonability, among \textit{Hearing} participants, the \textit{Bell} received an average score of 5.16 ($SD=1.40$) and \textit{Speech} received an average score of 5.60 ($SD = 1.32$). As for \textit{DHH} participants, the \textit{Bell} received an average score of 4.18 ($SD=1.94$) and \textit{Speech} received an average score of 5.27 ($SD = 1.74$).

We were interested in understanding whether participants value the \textit{Bell} and \textit{Speech} differently and whether there was a significant difference between \textit{Hearing} and \textit{DHH} participants; therefore, we explored the impact of Auditory Stimuli (\textit{Bell} and \textit{Speech}) and Group (\textit{Hearing} and \textit{DHH}) on the necessity and reasonability ratings. The NPVA revealed a significant main effect of \textit{Group} ($F=5.023, df=1, p=.025$) on necessity ratings. Post-hoc results confirmed that \textit{Hearing} participants ($M=5.36, SD=1.48$) gave higher necessity ratings than \textit{DHH} participants ($M=4.27, SD=1.91$). Regarding reasonability, the NPVA did not yield a significant difference.

\subsection{Ranking}
The ranking shows a preference for \textit{Speech} ($M=1.58, SD=0.69$, 19 ranked it first, while 13 ranked it second). It was followed by \textit{Bell} ($M=1.72, SD=0.66$, 14 ranked it first and 18 ranked it second), and \textit{Baseline} was rated the worst ($M=2.69, SD=0.62$, 28 ranked it as the third). \textit{Baseline} was largely disliked by \textit{Hearing} participants (only 1 participant did not rank it as the third); opinions among \textit{DHH} participants were varied. We noted that 3 \textit{DHH} participants ranked \textit{Baseline} as their first, 5 \textit{DHH} participants ranked it as second, and only 4 out of 11 ranked it as the third option.

Therefore, we were interested in understanding whether participants rank the \textit{Bell} and \textit{Speech} differently and whether there was a significant difference between \textit{Hearing} and \textit{DHH} participants. We followed the same analysis as described in the Necessity and Reasonability subsection. The NPVA revealed a significant main effect of \textit{Audio} ($F=7.551, df=1.959, p<.001$). Post-hoc pairwise comparisons suggested that \textit{Bell} ($M=1.72, SD=0.66$) and \textit{Speech} ($M=1.58, SD=0.69$) were ranked significantly better than \textit{Baseline} ($M=2.69, SD=0.62$, both $p<.001$). We also observed a significant interaction effect of Group and Auditory Stimuli ($F=5.079, df=1.959, p=.007$), post-hoc showed that the results of the main effect only held true for \textit{Hearing} participants (i.e., \textit{Baseline} was ranked significantly worse than \textit{Speech} [$p<.001$] and Bell [$p=.003$]).

\subsection{Qualitative Results}
The qualitative data were analysed using inductive thematic coding, which allowed themes to emerge from the participants' own descriptions. Two coders conducted the initial coding independently; Coder 1 designed the study and conducted the interviews, while Coder 2 was not involved in data collection. Afterward, both coders met to discuss, refine, and reconcile the themes to ensure consistency and reliability. We present anecdotal feedback and participants' opinions based on the specific questions asked during the interviews. To facilitate descriptive comparison, we counted the frequency of mentions for each theme across participants, providing an overview of both common and contrasting views while maintaining the qualitative depth of interpretation.

\subsubsection{Baseline} Slightly over half of the participants (N=20; Hearing: 14; DHH: 6) said they could hear clearly about the Baseline (AVAS and tire-pavement volume changes). Five participants (Hearing: 3; DHH: 2) mentioned they may have heard it, but they could not distinguish it from the background during the study. The remaining participants noted that they did not hear it. Several participants (N=25; Hearing: 17; DHH: 8) said there was an impact of the Background Noise, 8 explicitly mentioned that the louder background noise made it harder to hear and 1 mentioned that ``the quiet environment helped me focus on the visual'' [P7]. 

\subsubsection{Bell} Most participants (N=27; Hearing: 21; DHH: 6) said they could hear the Bell under both quiet and noisy backgrounds. In addition, 7 participants said they could only hear (1) when the vehicle is fully stopped (DHH: 1) and (2) in a quiet environment (Hearing: 4; DHH: 2). Among the 34 people who could hear the Bell, 24 participants (Hearing: 17; DHH: 7) agreed that the Background Noise impacted their perception of it. Among these 24 participants, 18 participants (Hearing: 13; DHH: 5) said louder background noise made it harder to hear. Interestingly, participants (N=4) mentioned that they felt the Bell sound blended in the background, making it difficult to hear. In addition, we observed that 3 DHH participants' hearing aid filtered the Bell sound as part of the Background Noise, especially in a noisy environment.

Overall, 9 participants explicitly praised the Bell, perceived it to be ``good'', ``nice'', and ``perfect''. This could be due to its ``audibility in quiet environment'' [P22], ``no language barrier'' [P10], and choice of timing and frequency being "good pace", ``well-timed pauses'' [P4, P6, P27]. 7 (Hearing: 6; DHH: 1) participants felt positive about the tone of the sound, described it to be ``pleasant'', but 3 others (Hearing: 2; DHH: 1) perceived it to be ``too calm'', ``too static'', and ``mismatch for AV context''.

Four participants (all Hearing) argued that they found the use of Bell familiar in real-life, with a similar pattern has been used by ``tram'', ``train level crossing'', and is ``similar to pedestrian crossing beeping''. However, this was opposed by 15 others (Hearing: 10; DHH: 5), who mentioned that the ``Bell has a weak association with traffic'' and ``its meaning is unclear''. For instance, P1 said, ``It sounds like a church bell, and I don't think it means it is about to slow, it doesn't link to it.'' P8 mentioned, ``I wouldn’t immediately associate the bell sound with a car''. We followed up by asking what if Bell would be used for auditory eHMI, they suggested ``it needs more sophisticated sound design to convey the message effectively'' [P16] or ``maybe use a beeping sound instead'' [P19]. 5 participants (all Hearing) said the volume is too low and 6 participants (Hearing: 4; DHH: 2) said it was too low in the noisy environment. 3 participants (Hearing: 2; DHH: 1) explicitly said the Bell had ``Blended into the background'' [P7, P15, P36] and either ``didn't stand out clearly'' [P15] or ``became indistinguishable'' [36].

\subsubsection{Speech} Most participants (N=30; Hearing: 25; DHH: 5) said they could hear the Speech eHMI under both noise levels. Unlike the Bell sound, all 25 Hearing participants were able to hear Speech. Additionally, 5 DHH participants were able to hear the Speech eHMI to some extent, either (1) an incomplete message (N=3), (2) some verbal sound but could not get what it said (N=1), or (3) clearly only under a quiet environment (N=1). Of the 35 participants who could hear or partly hear speech eHMI, 24 participants (Hearing: 17; DHH: 7) agreed that background noise impacted their ability to perceive speech eHMI. 23 out of 24 participants (Hearing: 16; DHH: 7) said louder noise made it harder. 

Overall, 12 participants (Hearing: 10; DHH: 2) explicitly praised the Speech, perceived it to be ``good'', ``best'', ``perfect''. This could be because Speech helped with understanding vehicle intention (N=11; Hearing: 8; DHH: 3), for instance, ``Hearing the speech gave clear identification that the vehicle was stopping'' [P17]. Additionally, it could also be due to the voice helping build the trust (N=2; Hearing: 1; DHH: 1). 14 participants (Hearing: 11; DHH: 3) liked the message content design as they were ``clear'', ``simple'', ``easily recognisable'', ``concise''. Despite it being simple, 2 participants felt ``odd'' and ``unnatural'' to hear that the vehicle used the phrase ``I am'' [P13, P18]. They said they would prefer alternative wording in the third person like ``the vehicle is''. There were also some suggestions about the issue of delivery: 5 participants (all Hearing) said the designed gap (i.e., 0.5 second) was not enough and would be better to have a larger gap.

Although the choice of robotic women's voice was liked (``pleasant tone'' and a ``good delivery'', ``it is distinct from human voices'') by 3 participants (Hearing: 2; DHH: 1), it also received criticism. P36 mentioned that ``a high-pitched robotic female voice would be difficult for people with high-frequency hearing loss to hear.'' To improve this, the suggestion of ``a deeper voice might be more accessible'' by P34 and P36 could be explored. In total, 7 participants (Hearing: 5; DHH: 2) said the volume was not loud enough, with 3 of them (Hearing: 2; DHH: 1) explicitly said in the loud setting, they felt ``hard to to hear'' [P15, P17] or ``only able to hear when it was close in noisy environment'' [P35]. 5 out of 7 commented that they could not hear when the vehicle was at a distance, which they understood later, as it could be 10 meters or even farther away.

\subsubsection{Real-life Sound Distraction}\label{real_life_sound}
Siren signals---including police, ambulance, and fire sirens, as well as fire alarms---are most frequently mentioned by the participants (N=16; Hearing: 11; DHH: 5). These sounds were perceived as critical and urgent; they could indicate a potential danger hazard and a signal of need to stop. Vehicle awareness-inducing sounds (i.e., standard engine hum, tyre noise, slow approaches, and reversing announcements) were mentioned by 9 participants (Hearing: 7; DHH: 2). These were helpful for general awareness but not urgent. Reversing announcements, in particular, were described positively for their clarity without being intrusive. Aggressive driving behaviours---such as car horns and screeching brakes---were reported by 9 participants (Hearing: 6; DHH: 3). They perceived these sounds as ``stressful'' or ``unpleasant'', and often linked them to ``reckless driving''. Similarly, show-off driving behaviours, including loud acceleration, racing, and modified exhausts, were mentioned by 8 participants (Hearing: 5; DHH: 3) and were rated as highly distracting as they stood out in the soundscape but were less about imminent safety. 6 participants (Hearing: 4; DHH: 2) mentioned environmental noise like construction, weather, and shouting as these sounds could mask more important auditory signals, forcing pedestrians to rely more heavily on visual cues. 

\subsubsection{Cross without Auditory Stimuli} Several participants (N=26; Hearing: 19; DHH: 7) said they would be able to cross without auditory eHMI, while 6 participants (Hearing: 4; DHH: 2) said visual eHMI alone is not enough, as they lacked trust in the use of visual eHMI alone. This statement was also mentioned by people who said they could cross with only visual eHMI, 11 out of 26 (Hearing: 10; DHH: 1) said auditory eHMI could help with ``confidence'', ``confirmation'', and create a ``safety layer if visuals are missed due to distraction'', therefore, they also explicitly expressed a preference for combined audio-visual cues.

\section{Discussion}
eHMIs have been researched for several years as a potential solution to improve communication between AVs and other road users. However, current research lacks focus in several critical areas: (1) evaluating these concepts with disabled pedestrians, particularly DHH people; (2) identifying accessibility barriers that auditory eHMIs may introduce; (3) understanding how DHH pedestrians interpret and interact with auditory cues, and ensuring that such cues do not inadvertently disadvantage them; and (4) examining how these concepts perform in realistic urban soundscapes, including how well different auditory eHMIs work under varying background noise conditions. Our study is the first VR simulation study that evaluates the effects of Background Noise on the perception of Auditory Stimuli among hearing and DHH people regarding their crossing experience (i.e., trust, acceptance, perceived safety, mental load) and behaviour (i.e., gaze behaviour, step-in road time, early step into the road count). 

\textbf{RQ1: How do the ratings for experience and behaviour differ between Hearing and DHH participants?} To conclude \textbf{RQ1}, our results showed some measurements of experience (i.e., usefulness rating among all conditions, safety rating among Bell and Speech auditory eHMI) was rated significantly higher by Hearing participants than DHH participants. The significantly higher usefulness ratings and safety ratings among Hearing people could be because DHH people did not well perceive the auditory eHMIs---our interview confirmed that all Hearing participants were able to perceive one of the auditory eHMIs under both conditions fully, and 21 out of 25 were able to perceive both auditory eHMIs under both background noise conditions fully; however, only 7 out of 11 DHH participants were able to fully perceive one of the auditory eHMIs under both conditions, and only 2 DHH participants were able to perceive both auditory eHMIs under both background noise conditions fully (see Table \ref{table:DHH_able_to_hear}). 

Based on our statistical results on necessity ratings, another explanation could be that auditory eHMIs may be significantly less necessary for most DHH participants than for Hearing participants. However, our interview also revealed that four DHH participants would not be able to cross without auditory stimuli, as they believed that it gave them confidence and made them feel safer (e.g., P30: "Audio gives confirmation and builds confidence in what I'm already perceiving visually, especially with cues like the bell." P32: "I find that a safer thing than hearing the bell or the voice", and P36: "I prefer having some sound to confirm that something is happening. It raises my situational awareness and makes me feel more confident.").

\begin{table*}[ht]
\caption{Results of DHH Participants regarding whether or not they were able to hear the designed auditory eHMI among quiet or noisy background (ordered based on severity of the hearing loss).}
\begin{tabular}{p{0.05\linewidth} p{0.1\linewidth} p{0.1\linewidth} p{0.075\linewidth} p{0.15\linewidth} p{0.07\linewidth} p{0.07\linewidth}p{0.07\linewidth}p{0.07\linewidth}}
\toprule
    ID & Left Ear \newline Hearing Loss & Right Ear\newline Hearing Loss & Identity & Preferred \newline Communication & Bell Quiet & Bell Loud & Speech Quiet & Speech Loud\\
    \midrule
    P31 &	Moderate    & Severe      & HoH   & English &	Yes&	Yes&	Yes&	Yes\\
    P32 &	Moderate    & Severe      & deaf  & English &	Yes&	Yes&	Yes&	No\\
    P26 &   Severe      & Severe      & HoH   & English &	Yes&	Yes&	Yes&	No\\
    P29 &	Severe      & Severe      & HoH   & English &	No&	No&	Yes&	No\\
    P34	&   Severe      & Severe      & deaf  & English &	Yes&	No&	Yes&	Yes\\
    P28 &	Profound    & Severe      & HoH   & English &	Yes&	No&	Yes&	Yes\\
    P36 &	Severe	    & Profound    & deaf  & English &   Yes&	Yes&	Yes&	No\\
    P30 &	Profound	& Profound    & deaf  & English &   Yes&	Yes&	Yes&	No\\
    P25 &   Profound    & Profound    & Deaf  & BSL     &   Yes&	No&	Yes&	Yes\\
    P27	&   Profound    & Profound    & Deaf  & BSL     &   Yes&	Yes&	Yes&	Yes\\
    P33 &	Profound	& Profound    & deaf  & BSL     &	No&	No&	No&	No\\
    \bottomrule
    \end{tabular}
  \label{table:DHH_able_to_hear}
  \Description{Table shows demographic of deaf and hard of hearing participants, with these columns show data from left to right to be: ID, left ear hearing loss level, right ear hearing loss level, identity, preferred Communication, and whether or not clearly heard the proposed (1) bell eHMI in a quiet background environment, (2) bell eHMI in a loud background environment, (3) speech eHMI in a quiet background environment, (4) speech eHMI in a loud background environment.}
\end{table*}

\textbf{RQ2: What impact does the Background Noise have on pedestrians regarding experience and behaviour?} To answer \textbf{RQ2}, we found that loud background noise impaired participants' crossing experience (trust, usefulness, safety, mental workload) but had no impact on the crossing behaviour. This finding corroborates the literature that loud background noise also has a negative impact on road crossing, in line with other daily activities such as office tasks \citep{BanburyBerry1998Disruption}. Early studies \citep{TAPIRO2018219} suggest that background noise caused participants to choose smaller crossing gaps, take more time to make crossing decisions, and be slower to respond to the crossing opportunity; these were not found in our work. A possible explanation could be that the background noise used in our study was more constant, there was no sudden and momentary sound that caused a sudden change in sound volume and frequency. \citet{TAPIRO2018219} employed sounds that are sudden and momentary (e.g., shop alarm, cyclist passing, noisy siren), which were also found to be distracting to our participants based on our qualitative data. Future exploration on sudden and momentary sounds are needed.

Early studies on the effect of noise on gaze behaviour during conversation showed people may show increased attention to the mouth to compensate for ambiguous auditory input \citep{Hadley2019}. We hypothesised that participants might allocate more visual attention to active eHMI components under loud background noise conditions to compensate for hearing ambiguous audio. However, we found no significant differences between background noise conditions, indicating that gaze behaviour (duration and percentage) remains stable across different background conditions. This could be because road crossing is much more visually demanding compared to speech-focused conversation \citep{SOARES2021202,PUGLIESE2020105344,app10082913} where participants had already paid more attention to the vehicle and the active eHMIs. It could also be because there is less need for compensation via gaze shifts due to the designed auditory eHMI being concise, repeated, and predictable once heard.

\textbf{RQ3: What impact do the Auditory Stimuli of the Audio-Visual eHMI have on pedestrians regarding experience and behaviour?} 
Our results showed that additional auditory eHMIs like Bell and Speech would improve experience (trust, usefulness, safety). This finding supports the auditory eHMI literature conducted in video-based research~\citep{multi_modal_dey}, VR simulation research~\citep{Mark_Vision}, and real-world Wizard-of-Oz research~\citep{BINDSCHADEL202359}. However, we did not observe a significant improvement in providing additional auditory eHMIs on the pedestrian’s crossing behaviour, such as eye gaze behaviour and step into the road decision making. This does not support the real-world Wizard-of-Oz research~\citep{BINDSCHADEL202359}, where participants made faster crossing decisions when the intention of the vehicle was played. To answer \textbf{RQ3}, providing additional auditory eHMIs, such as Bell and Speech, improves experience but does not impact behaviours. 

\textbf{Multi-modal eHMIs}. Our participants and related works~\citep{multi_modal_dey,Mark_Vision} highlighted the benefits of multi-modal eHMI. In particular, our findings support the use of audio-visual eHMI. However, the auditory eHMI requires careful consideration to ensure it supports rather than overwhelms pedestrians. Transport noise already ranks among Europe’s top three environmental health threats, with more than 20\% of Europeans exposed to harmful levels~\citep{EEA2025Noise}. Thus, auditory eHMI should be designed in line with the WHO’s environmental noise guidelines~\citep{WHO2018NoiseGuidelines}, while also ensuring that signals are both perceptible under varying background noise environments~\citep{10.1145/3409120.3410646}. It is worth noting that auditory eHMI may still fail to work for all people, e.g., P33 cannot hear all designed auditory eHMI even with the help of hearing aids who claimed during the interview that he could not access truck reverse sound as well in the real-life, which is typically loud (112 dB) and with a high frequency~\citep{UNECE_WP29_2022_88E}. However, P33 agreed that auditory eHMI would still be useful for hearing people (e.g., "I think the speech is important for hearing people" but then said, because now he has become deaf (due to an accident), "it doesn’t matter") and said he would be able to cross with purely visual eHMI. Future work could explore other modalities, such as haptics, to enable multi-modal eHMIs for people who were unable to access auditory eHMIs.

\textbf{Auditory Stimuli and Hearing Technologies}. Statistically, we could not conclude which auditory eHMI (Speech or Bell) was better, as we did not find any significant difference between them among all measurements we had (i.e., experience, behaviour, necessity, reasonability, and ranking). This could be because both auditory eHMIs communicated the intention at the same starting time (i.e., 1.5 seconds after the vehicle initiated the yielding), as prior work suggests the timing of information is more important \citep{10.1145/3568162.3576979}. We observed several concerns regarding each method during the interview. For Bell eHMI, (1) it is hard to understand the meaning/intention of the Bell at the beginning, (2) it has a weak association with traffic, and (3) there is a risk Bell would be masked with the background sound (potentially filtered by their hearing aids as described by participants but also supported by hearing aids review \citep{Launer2016}). 

As for Speech eHMI, a significant concern observed is that when DHH participants focused on crossing, they may only partially hear the message content (i.e., "Stopped" from "I'm Stopped"), which led to confusion. Early study \citep{Mark_Vision} compared a low content message ("Cross") and a high content message ("I’m stopping, you can cross") and found that the high content message could reduce cognitive load for low vision or blind people. The high content message design could cause more issues for DHH people, as there is a greater risk of missing critical details. Researchers and designers should consider working with DHH people and different hearing technologies to amplify auditory eHMI designs rather than filter out \citep{moore2019design}.

\textbf{Variability within the DHH Group among Auditory eHMI}. The DHH group exhibited greater variability than the hearing group. A key reason for this could be the different audibility of the auditory eHMIs in the presence of loud background noise. As shown in Table \ref{table:DHH_able_to_hear}, DHH participants tend to find it easier to access auditory eHMIs in quiet environments, while the access to each eHMI under loud background environments varied from person to person (and not solely due to hearing loss level): some could hear the Speech eHMI but not the Bell eHMI (i.e., P25, P28, P34), while others perceived the opposite (i.e., P26, P30, P32, P36). The hearing group demonstrated a much more consistent trend. All hearing participants were able to hear the Speech eHMI in both quiet and noisy backgrounds (although louder noise required more effort). They were also able to hear Bell eHMI under quiet backgrounds, with only 4 out of 25 participants being unable to detect the Bell eHMI in the loud condition (compared to DHH participants, this is a much smaller percentage). 

Another reason could be the hearing technology being used. Although everyone in our study uses hearing technologies, each device may have different functions and priorities (e.g., amplify speech, filter background noise, or a mixture of both \citep{Plomp1994,Kates2002}). Future research with DHH people should make sure the details of the hearing technology brand, model, and key functions are collected for further analysis and record what functions are activated. 

\subsection{Practical Implications}
Limited work explored the use of eHMI with disabled people \citep{Mark_Vision, Asha_wheelchair,10.1145/3546717}, we \textbf{call for further research to involve disabled people} as we explored differences in crossing experiences between hearing and DHH people. Prior works with hearing eHMIs were usually tested in environments with limited to no soundscape, where participants had no issues assessing the auditory eHMI \citep{Mark_Vision,multi_modal_dey}. We found that loud background noise had a negative impact on participants' perception of the crossing experience and sometimes led to participants being unable to hear the auditory eHMI. Therefore, we suggest \textbf{further work in eHMIs should employ background noise} to ensure the auditory eHMIs would work under different environments (e.g., quiet rural area, quiet urban area, busy urban area). 

We recommend that eHMI design and research should \textbf{enable audio-visual (multi-modal) eHMI} so that when pedestrians face situations where one modality was missed, they could still rely on the other. The reason for enabling audio-visual (multi-modal) eHMIs, such as the combination used in our study (Visual: Abstract Light + Text, Audio: Text or Bell), was that combined eHMIs significantly improve the crossing experience compared to just the visual eHMI and sound stimuli generated from the driving vehicle. Regarding which auditory eHMI would be the best, there is not enough evidence from our results to conclusively answer this question. Further research on auditory eHMIs is needed, and we suggest that such research focus on \textbf{creating hearing technologies-friendly auditory eHMI}. Based on our qualitative findings, some hearing technologies could mask and filter the Bell as part of the background noise. Although speech can be amplified, it could also cause issues for users who could not capture the entire phrase (i.e., a DHH participant only heard ``Stop'' from the phrase ``I'm stopped''); which points to topics that need to be addressed in future work. This could be done by working with hearing technology companies to (1) ensure that eHMI messages are clearly delivered or (2) filter them if they are not clearly captured to avoid confusing the participants.

\subsection{Limitation and Future Work}
This research has some limitations, which can also serve as directions for future studies. As limited work has been conducted on auditory eHMI and especially with DHH people, we employed a simple environment with a single controlled scenario featuring a non-signalised crossing to remove most distractions and allow participants to concentrate on the features~\citep{DEB2018135,Mark_Vision}. This controlled setup prioritised internal validity by enabling control over key factors; however, while appropriate for this purpose, it may limit ecological validity~\citep{scalability_Mark}. Future work could explore the scenario where participants need to cross (1) with other pedestrians~\citep{scalability_Mark}, (2) in mixed traffic with both manual and automated vehicles~\citep{10.1145/3409120.3410646}, or (3) at a controlled traffic or zebra crossing to explore how they might affect pedestrians’ crossing experience and behaviours. Ultimately, exploring the eHMI in field studies to maximise the ecological validity~\citep{10.1145/3342197.3345320}.

The sample size for DHH people (N=11) is relatively small due to the difficulty in finding disabled participants and is uneven compared to the hearing group (N=25). The future work should include more participants and an even number of samples across the groups. In addition, low vision or blind people could heavily rely on auditory eHMI. Exploring other disabled groups in the future will enhance the overall comprehensiveness of our findings. Our study only involved a single experimental session and a single country. Future research could adopt a longitudinal design to test the findings in different countries and cultures, determining whether they are applicable in the long term across various contexts \citep{10.1145/3699778}.

As for the background noise perspective, we only explored background noise with good representations of standard urban environments (confirmed to be realistic by our participants in both the iterative testing and the formal study). We did not include typical distracting sounds, such as sirens, aggressive driving sounds, and sounds induced by show-off driving behaviours, as mentioned in Section \ref{real_life_sound}. This is because these sounds also involve visual distractions that add complexity to the scenario design, while we want to focus on the background noise as a more controlled starting point. Future work could explore how these sounds would impact DHH people's crossing experience with auditory eHMIs. Regarding auditory eHMI, we only utilised existing ones from prior work \citep{multi_modal_dey} for our studies; we did not employ a comprehensive sound creation process to explore all the characteristics of auditory eHMI. Additionally, the interval between each auditory eHMI would also impact pedestrians' experience. As we discussed in \textbf{Auditory Stimuli and Hearing Technologies}, there is also a need for further in-depth investigation on auditory eHMI design.  

\section{Conclusion}
Through a VR simulation, this research investigated the effect of background noise (quiet and loud) with auditory stimuli (baseline, bell, speech) for AV-pedestrian communications. We also explored the crossing experience and crossing behaviours between Hearing participants (N=25) and DHH participants (N=11) with the intention of understanding how we can better design eHMI for DHH people. Our results draw three conclusions based on the crossing scenario we evaluated: (1) Auditory stimuli should be carefully designed with consideration for DHH people and ensure that those auditory stimuli can be captured by their hearing technologies, so that their experience and behaviour are not impaired. (2) Loud background noise level would significantly impact pedestrians' crossing experience, but not the crossing behaviour we measured. (3) Providing additional auditory stimuli (i.e., Bell or Speech) could improve crossing experience, but has no impact on crossing behaviour. We also proposed four practical implications that pave the way for inclusive eHMI design and research.

\begin{acks}
The authors thank all participants for their time. This work was funded by the Royal Society (RG\textbackslash R1\textbackslash 241114).
\end{acks}

\bibliographystyle{ACM-Reference-Format}
\bibliography{sample-base}

\end{document}